\documentclass[12pt,twoside]{article}
\usepackage{amssymb}
\usepackage{amsmath,amsthm}
\usepackage{latexsym}

\begin{document}
\begin{center}
{\Large {\bf Kinetic Equations for Quantum Many-Particle
Systems\bigskip\bigskip\\}}
{\large{Herbert Spohn}}\footnote{{\tt spohn@ma.tum.de}}\medskip\\
Zentrum Mathematik and Physik Department, TU M\"{u}nchen,\\
D - 85747 Garching, Boltzmannstr. 3, Germany\\
\bigskip\bigskip\bigskip\bigskip
{\large Modern Encyclopedia of Mathematical Physics\\
 Springer Selecta}
 \end{center}
\newpage
\section{The framework}\label{sec.1}

Kinetic equations are a widely used and highly successful method to
understand dynamical aspects of quantum many particle systems. As a
standard reference I mention the book by Ziman \cite{Z67} which in
large parts is based on the kinetic method.

My article deals with the issue to derive such an equation in a
well-defined limit from quantum dynamics. For sake of a simple
presentation I restrict myself to a weak interaction potential.
There is a parallel discussion for low density for which I have to
point to the literature \cite{S94,BCEP06}.
Let me first describe the general framework in admittedly rather
vague terms. The broad picture will be made precise in the remainder
of the article, where also the border between proven and conjectured
will be traced out.

We consider (nonrelativistic) quantum particles moving in three
dimensions and interacting through a short range potential. The
hamiltonian is of the generic form
\begin{equation}\label{1.1}
H=H_0+\sqrt{\varepsilon}H_\mathrm{I}\,,
\end{equation}
where $H_0$ denotes the free hamiltonian. $H_\mathrm{I}$ is the interaction
which is multiplied by a dimensionless strength parameter
$\sqrt{\varepsilon}$. We want to study a system of many particles
and, first, consider a spatially homogeneous situation, which is
idealized as an infinitely extended system in a spatially
translation invariant state. Amongst the many conceivable states
there is a special class of states characterized by being quasifree,
translation invariant, and time-invariant under the dynamics
generated by $H_0$. Such a state is uniquely characterized by a mass
density $\rho$, $\rho>0$, and a momentum distribution $W(k)/\rho$,
$\int dk W(k)=\rho$.

Under the full dynamics generated by $H$ such a quasifree state
changes in time but remains
translation invariant. The change is slow, since
the interaction is weak. By definition,  the kinetic time scale is the shortest scale
over which the
change in the momentum distribution is of order 1. Since transition
rates are roughly proportional to $|\sqrt{\varepsilon}H_\mathrm{I}|^2$, the
kinetic time scale should be $\varepsilon^{-1}t$, in microscopic
time units. On the kinetic time scale the momentum distribution
evolves according to an autonomous nonlinear transport equation.
Generically, on the basis of this equation, the long time limit will
be the thermal equilibrium momentum distribution for $H_0$, i.e.
for the noninteracting particles. Moreover, the kinetic equation
makes a quantitative prediction on how thermal equilibrium is
approached.

At this stage the difficulty of a proof becomes already apparent.
Since $\varepsilon\ll 1$, Equation (\ref{1.1}) looks like a standard
perturbation problem. But this perturbation  has to be controlled over
the very long times $\varepsilon^{-1}t$.

What happens if the initial state is not quasifree but still translation
invariant? This situation is analysed
in \cite{HL97} for a weakly interacting Fermi gas on a lattice. If the
initial state has good spatial clustering, then, under the $H_0$-dynamics
for long microscopic times, the state becomes quasifree without yet
changing the momentum distribution. The resulting  dynamical picture is
rather compelling: For ``reasonable'' initial states, the
time-evolved state becomes rapidly quasifree. At kinetic times the
interaction makes itself felt, changing the momentum  distribution
but maintaining the quasifree property. At even longer times the
system presumably tries to establish the equilibrium state for $H$,
which is not strictly quasifree. For fixed small $\varepsilon$,
physically the time scales are not perfectly separated. But
mathematically, a separation can be achieved through taking limits. The
first regime is $\varepsilon=0$ and $t\to\infty$. The kinetic
regime is $\varepsilon\to 0$ and $t\to\infty$, $\varepsilon t=\tau$
fixed, while the third regime is $\varepsilon$ fixed and small,
while $t\to\infty$. Needless to remark that the third regime is
inaccessible mathematically.

Usually, one also wants to include spatial variations, which however
are slow on the scale of the typical distance between particles. For
such slow variations, the $H_0$-dynamics can be approximated by
semiclassics, which means that particles move independently with
classical kinetic energy $\omega(k)$, $\omega$ being the dispersion
relation for the quantum particles. The kinetic time scale has been
determined already. To have the $H_0$-dynamics of the same order as
the collisions requires then a spatial variation of order
$\varepsilon^{-1}$ in microscopic units.
The momentum distribution varies now spatially on the kinetic scale.
Thus $W(k)$ is to be replaced by $W(r,k)$. The collision term is
strictly local in $r$, while the kinetic equation picks up the flow
term $-\nabla_k\omega(k)\cdot\nabla_r W(r,k)$ from the semiclassical
approximation. Close to $r$ the microscopic state is still quasifree
and translation invariant, at least in approximation.

\section{Model hamiltonians}\label{sec.2}
\setcounter{equation}{0}

There is some freedom in the choice of the microscopic model.
Firstly, the position space is either $\mathbb{R}^3$ or some Bravais
lattice which we take to be $\mathbb{Z}^3$. (We will comment on the
spatial dimension below.) The average number of particles is finite but
diverges as $\varepsilon^{-3}$ in the kinetic limit. Thus the
Hilbert space for the system is the boson Fock space
$\mathcal{F}_+$, resp.~fermion Fock space $\mathcal{F}_-$, over
either $L^2(\mathbb{R}^3)$ or $l_2(\mathbb{Z}^3)$. Let $a(x)$,
$a(x)^\ast$, $x\in\mathbb{R}^3$, resp.~$x\in\mathbb{Z}^3$, denote
annihilation and creation operators for the Bose, resp.~Fermi,
fields. Setting $\hbar=1$, they satisfy the commutation relations
\begin{equation}\label{2.1}
[a(x),a(x')]_{\mp}=0=[a(x)^\ast,a(x')^\ast]_{\mp}\,,
\end{equation}
\begin{equation}\label{2.2}
[a(x),a(x')^\ast]_{\mp}=\delta(x-x')\,,
\end{equation}
where $[\cdot,\cdot]_-$ denotes the commutator, $[\cdot,\cdot]_+$
the anticommutater, and $\delta(x-x')$ stands for the Kronecker
$\delta$ in case of $\mathbb{Z}^3$. The momentum space operators are
distinguished only through their argument and given by
\begin{equation}\label{2.3}
a(k)=(2\pi)^{-3/2}\int_{\mathbb{R}^3}d x e^{-ik\cdot x}a(x)\,,\quad
k\in\mathbb{R}^3\,,
\end{equation}
\begin{equation}\label{2.4}
a(k)=\sum_{x\in\mathbb{Z}^3} e^{-i 2\pi k\cdot x}a(x)\,,\quad
k\in[-\frac{1}{2},\frac{1}{2}]^3=\mathbb{T}^3\,.
\end{equation}\medskip
\textit{(i) Independent particles with random impurities} (Anderson
model). Particles hop to nearest neighbor sites and are subject to a
static potential $V(x)$. Then the hamiltonian reads
\begin{equation}\label{2.5}
H=-\frac{1}{2}\sum_{x\in\mathbb{Z}^3} a(x)^\ast(\Delta
a)(x)+\sqrt{\varepsilon}\sum_{x\in\mathbb{Z}^3} V(x)a(x)^\ast a(x)\,.
\end{equation}
Here $\Delta$ is the lattice Laplacian and
$\{V(x),x\in\mathbb{Z}^3\}$ is a collection of independent random
variables. The random potential mimics the interaction. The kinetic
equation will be linear, however.\medskip\\
\textit{(ii) Electron-phonon systems.} $a(x)$ is the Fermi field of the electrons and
$c(x)$ is the Bose field of the phonons. The hamiltonian is given by
\begin{eqnarray}\label{2.6}
&&\hspace{-36pt}H=-\frac{1}{2}\int_{\mathbb{R}^3}dx a(x)^\ast\Delta
a(x)+\int_{\mathbb{R}^3}dk
\omega(k)c(k)^\ast c(k)\nonumber\\
&&\hspace{36pt}+\sqrt{\varepsilon}\int_{\mathbb{R}^3}dx
\int_{\mathbb{R}^3}dx' a^\ast(x)a(x)g(x-x')\big(c(x')+c(x')^\ast\big)
\end{eqnarray}
with $\omega$ the phonon dispersion and $g$ a suitable coupling
function. The number of electrons is conserved. Assuming the phonon
system to be infinitely extended and in thermal equilibrium, the dynamics of
a single electron is governed by
a linear kinetic equation.\medskip\\
\textit{(iii) Weakly interacting fermions and bosons.} If the pair
potential is denoted by $V$, the hamiltonian reads
\begin{equation}\label{2.7}
H=-\frac{1}{2}\int_{\mathbb{R}^3}dx a(x)^\ast\Delta
a(x)+\int_{\mathbb{R}^6}dx dx' a(x)^\ast a(x')^\ast V(x-x')
a(x)a(x')\,.\medskip
\end{equation}
\textit{(iv) Phonons.} Phonons are bosons. For effective single band
models the free part is of the form
\begin{equation}\label{2.8}
H=\int_{\mathbb{T}^3}dk\omega(k)a(k)^\ast a(k)\,.
\end{equation}
$\omega$ is the dispersion relation. $\omega^2$ is the Fourier
transform of a rapidly decaying lattice function. For an acoustic
band $\omega(k)=|k|$ for small $k$, while for optical bands
$\omega(k)\geq \omega_0>0$ for all $k\in\mathbb{T}^3$. Expanding
around the harmonic potentials yields cubic and quartic
nonlinearities. For example, a quartic on-site potential would be
$\lambda\sum_{x\in\mathbb{Z}^3}(q_x)^4$, $q_x$ the scalar
displacement at $x$ away from the equilibrium position, while cubic
nonlinear springs correspond to
$\frac{1}{3}\sum_{x,y\in\mathbb{Z}^3}\alpha(x-y)(q_x-q_y)^3$. The
respective interaction hamiltonians are
\begin{equation}\label{2.9}
H_{\mathrm{I}4}=\lambda\int_{\mathbb{T}^{12}}dk_1 dk_2 dk_3 dk_4
\delta(k_1+k_2+k_3+k_4)\prod^4_{j=1}\big(2\omega(k_j)\big)^{-1/2}
\big(a(k_j)+a(-k_j)^\ast\big)
\end{equation}
and
\begin{eqnarray}\label{2.10}
&&\hspace{-36pt}H_{\mathrm{I}3}=\frac{1}{3}\int_{\mathbb{T}^9} dk_1 dk_2 dk_3
\delta(k_1+k_2+k_3)\nonumber\\
&&\hspace{-2pt}\sum_{x\in\mathbb{Z}^3}\alpha(x)
\prod^4_{j=1}\big(2\omega(k_j)\big)^{-1/2}\big(e^{i2\pi k_j\cdot
x}-1\big)\big(a(k_j)+a(-k_j)^\ast\big)\,.
\end{eqnarray}
$H_0+\sqrt{\varepsilon}H_{\mathrm{I}3}$ is not bounded from below, a defect
not  yet seen on the kinetic time scale. For phonon systems the
particle number is not conserved. However, it is possible that on
the kinetic level only processes with number conservation are
allowed.

\section{Initial state}\label{sec.3}
\setcounter{equation}{0}

As explained in Section \ref{sec.1}, for the spatially homogeneous
case the initial state is translation invariant, quasifree, and
invariant under the $H_0$ dynamics. Therefore $\langle
a(k)\rangle=0=\langle a(k)^\ast\rangle$ and
\begin{equation}\label{3.1}
\langle a(k')^\ast a(k)\rangle = W(k)\delta(k-k')\,,
\end{equation}
\begin{equation}\label{3.2}
\langle a(k')^\ast a(k)^\ast\rangle =0\,,\;\langle
a(k')a(k)\rangle=0\,.
\end{equation}
Here $W(k)\geq 0$ and for fermions in addition $W(k)\leq 1$. All other moments
are determined through the two-point function, i.e.
\begin{equation}\label{3.3}
\langle \prod^m_{j=1}a(k_j)^\ast \prod^m_{j=1}a(k'_j)\rangle =\,
\begin{matrix}{\textrm{per}} \\ {\det}\end{matrix}\;\big\{\langle a(k_i)^\ast
a(k_j)\rangle\big\}_{i,j=1,\ldots,m}
\end{equation}
with all other mononomials having vanishing expectation. Such a state has
infinite energy and is not supported on Fock space.

To model the spatially inhomogeneous situation, we give ourselves a
spatially dependent Wigner function $W(r,k)$, $W(r,k)\geq 0$,
$W(r,k)\leq 1$ for fermions, and of rapid decay, in particular $\int
dr \int dk W(r,k)<\infty$. Secondly we choose a sequence of trace
class operators, $R^\varepsilon$, $\varepsilon>0$, on $\mathcal{F}$,
$R^\varepsilon\geq 0$ and $R^\varepsilon\leq 1$ for fermions. Let
$R^\varepsilon(x,y)$ be the corresponding integral kernel. Then the
sequence of initial states, $\langle\cdot\rangle_\varepsilon$,
$\varepsilon>0$, still satisfies (\ref{3.1}), (\ref{3.2}), and
(\ref{3.3}) with
\begin{equation}\label{3.4}
\langle a(x)^\ast a(y)\rangle_\varepsilon = R^\varepsilon(x,y)\,.
\end{equation}
The Wigner function $W$ appears through the limit $\varepsilon\to 0$
by requiring
\begin{equation}\label{3.5}
\lim_{\varepsilon\to 0}
R^\varepsilon\big(\varepsilon^{-1}r+\frac{1}{2}\eta,
\varepsilon^{-1}r-\frac{1}{2}\eta\big)=\int_{\mathbb{R}^3}dk
e^{ik\cdot\eta}W(r,k)\,.
\end{equation}
In particular, for the number of particles, $N$, one has
\begin{equation}\label{3.6}
\langle N\rangle_\varepsilon\simeq \varepsilon^{-3}\,.
\end{equation}

(\ref{3.5}) implies that at $r$, on the kinetic scale, in the limit
$\varepsilon\to 0$ the state is quasifree with correlator
$W(r,k)\delta(k-k')$ ($r$ is fixed here). Spatial averages on the
kinetic scale are self-averaging (law of large numbers). To see
this, let $A$ be a local observable and $\tau_x$ the shift by $x$.
Spatial average on the kinetic scale means
\begin{equation}\label{3.7}
\varepsilon^3\int_{\mathbb{R}^3}dx \chi_\Lambda(\varepsilon x)\tau_x
A=n^\varepsilon(\Lambda,A)
\end{equation}
with $\chi_\Lambda$ the indicator function of a fixed box $\Lambda$.
Then
\begin{equation}\label{3.8}
\lim_{\varepsilon\to
0}n^\varepsilon(\Lambda,A)=\int_{\mathbb{R}^3}dr\chi_\Lambda(r)\langle
A\rangle_{W(r,\cdot)}
\end{equation}
in the sense that the variance of $n^\varepsilon(\Lambda,A)$
vanishes as $\varepsilon\to 0$. $\langle A\rangle_{W(r,\cdot)}$ is
here the average of $A$ in the translation invariant, quasifree
state with correlator
\begin{equation}\label{3.9}
\langle a(k')^\ast a(k)\rangle_{W(r,\cdot)} =W(r,k)\delta(k-k')\,.
\end{equation}

\section{Kinetic limit}\label{sec.4}
\setcounter{equation}{0}

Let $U^\varepsilon(t)=\exp[i(H_0 +\sqrt{\varepsilon}H_{\mathrm{I}})t]$ be the
unitary group generated by $H$. (In brackets we note that in all
models $H$ is self-adjoint with domain $D(H_0)$ under reasonable
assumptions on the coefficients, except for
$H_0+\sqrt{\varepsilon}H_{{\mathrm{I}}3}$ which has to be stabilized by adding
a small quartic potential.) As initial state we choose the sequence
$\langle\cdot\rangle_\varepsilon$ of quasifree states from Section
\ref{sec.3}. It is time evolved to $\langle\cdot\rangle_\varepsilon
(t)$ according to $\langle A\rangle_\varepsilon (t)=\langle
U^\varepsilon(t)^\ast A U^\varepsilon(t)\rangle_\varepsilon$ for all
$A\in B(\mathcal{F})$. To prove the kinetic limit firstly means to
establish the existence of
\begin{equation}\label{4.1}
\lim_{\varepsilon\to 0}\langle
a(\varepsilon^{-1}r-\frac{1}{2}\eta)^\ast
a(\varepsilon^{-1}r+\frac{1}{2}\eta)\rangle_\varepsilon(\varepsilon^{-1}t)
=\int_{\mathbb{R}^3}dk e^{ik\cdot\eta}W(r,k,t)\,.
\end{equation}
(For model \textit{(i)} one averages also over the random potential. For
model \textit{(ii)} one considers only a single electron and averages over
the thermal state of the phonons.) Clearly an issue is the sense in
which the limit (\ref{4.1}) is intended. Usually one proves weak
convergence, but pointwise convergence could hold. More ambitiously
one can study higher moments which should follow the pattern
explained in Section \ref{sec.3}. Only the Wigner function $W(r,k)$
is to be replaced by the time-evolved Wigner function $W(r,k,t)$. A further item would be
the law of large numbers and its central limit corrections.

Secondly the time evolution of the Wigner function $W(r,k,t)$ must
be governed by an autonomous transport equation. If $H_{\mathrm{I}}=0$,
such a property and the
limit (\ref{4.1}) are easily established with the
result
\begin{equation}\label{4.2}
\frac{\partial}{\partial t}W(r,k,t) + \nabla_k\omega(k)\cdot\nabla_r
W(r,k,t)=0\,.
\end{equation}
If $H_{\mathrm{I}}\neq 0$, the transport equation (\ref{4.2}) acquires a
collision term as
\begin{equation}\label{4.3}
\frac{\partial}{\partial t}W(r,k,t) + \nabla_k\omega(k)\cdot\nabla_r
W(r,k,t)=\mathcal{C}(W)(r,k,t)\,.
\end{equation}
The collision operator $\mathcal{C}$ is local in space-time and we
only have to display its action on the momentum variable. Of course
$\mathcal{C}$ depends on the particular model. Let us give only two
examples. For an electron in a random potential one obtains
\begin{equation}\label{4.4}
\mathcal{C}(W)(k)=2\pi\int_{\mathbb{R}^3} dk'\widehat{\vartheta}(k-k')
\delta\big(\omega(k)-\omega(k')\big)\big(W(k')-W(k)\big).
\end{equation}
Here $\omega$ is the dispersion relation for the electron and 
$\widehat{\vartheta}$ is the Fourier transform of the correlator of the random potential,
$\vartheta(x-x') = \langle V(x)V(x')\rangle$, $\langle V(x)\rangle = 0$. The collisions uniformize 
the momentum over the shell of constant energy. Thus
with this collision term Eq.~(\ref{4.3}) has the property that as
$t\to\infty$ the distribution function of the electron converges to $W(k)
= \delta(\omega(k) - E)$, given the initial data have kinetic energy $E$.

Our second example is the Uehling-Uhlenbeck equation for weakly
interacting quantum fluids \cite{UU33}, example \textit{(iii)}, which for
fermions was first derived by Nordheim \cite{N29}. The collision term reads
\begin{eqnarray}\label{4.5}
&&\hspace{-50pt}\mathcal{C}(W)(k_1)= \int dk_2 dk_3 dk_4\Phi(k_1,k_2|k_3,k_4)\nonumber\\
&&\hspace{15pt}[W_3 W_4(1+\theta W_1)(1+\theta W_2)-W_1W_2(1+\theta
W_3)(1+\theta W_4)]\,,
\end{eqnarray}
$W_j=W(k_j)$, with $\theta=1$ for bosons and $\theta=-1$ for fermions.
The collision rate encodes energy-momentum
conservation and the interaction potential in the Born approximation as
\begin{eqnarray}\label{4.6}
&&\hspace{-40pt}\Phi(k_1,k_2|k_3,k_4)=|\widehat{V}(k_1-k_3)+
\theta \widehat{V}(k_2-k_3)|^2 \delta(k_1+k_2-k_3-k_4)\nonumber\\
&&\hspace{52pt}\delta(\omega_1+\omega_2-\omega_3-\omega_4)\,,\quad
\omega_j=\frac{1}{2}k_j^2\,.
\end{eqnarray}
Note that (\ref{4.5}) is in fact cubic. The added term makes
symmetry of the collision operator more transparent. Phonon
collision operators are listed in \cite{S06}. They were first derived by Peierls \cite{P29}.

One part of the limit theorem must ensure the existence of a unique
solution to the transport equation (\ref{4.3}). For a linear
collision operator tools are available. For a nonlinear collision
operator local in time existence can always be established. To have
a solution global in time is a difficult issue. Since collisions are
at some particular  spatial location, the nonlinearity is well-defined only if
$\sup_{r,k}W(r,k,t)$ is bounded, a property one does not know how to
establish in generality. An exception are fermions governed by
(\ref{4.5}) with $\theta=-1$. Then the solution must satisfy $0\leq
W(r,k,t)\leq 1$ for all $t$, see \cite{D94} for a complete
discussion.

\section{Mathematical results}\label{sec.5}
\setcounter{equation}{0}

For a classical system of hard spheres at low density all facets of
the kinetic limit are proved under the sole restriction of
kinetically short times, i.e.~$|t|\leq t_0=1/5$-th of the mean free
time  \cite{L75}, see also \cite{CIP94,S91}. The issue under discussion is to carry out a
corresponding program for weakly nonlinear wave equations,
resp.~linear wave equations with weak disorder.

The zeroth order step is to have a convincing formal derivation and
thereby to determine the collision operator. Such formal derivations
abound in the physics literature. They span from a direct application
of Fermi's golden rule to sophisticated diagrammatic expansions. To
me the most convincing argument is a truncation procedure based on
the assumption that the state at microscopic time
$\varepsilon^{-1}t$ is approximately quasifree. The details can be
found in \cite{ESY04}, see also \cite{BCEP04}, for weakly interacting quantum fluids and in
\cite{S06} for weakly interacting phonons.

Up to now, any proof is based on the time-dependent perturbation
theory, which means to expand
\begin{equation}\label{5.1}
e^{-iHt}=e^{-iH_0t}+\sum^\infty_{n=1}\int_{0\leq t_1\leq\ldots\leq
t_n\leq t}dt_1\ldots dt_n e^{-iH_0(t-t_n)}V_{\mathrm{I}}\ldots V_{\mathrm{I}}
e^{-iH_0t_1}\,.
\end{equation}
Inserting in $\langle U(t)^\ast a(y)^\ast
a(x)U(t)\rangle_\varepsilon$ and working out the average over the
initial state (resp.~over the random potential)  this
expectation is then represented as a sum of high-dimensional oscillatory integrals (=
diagrams). Very crudely the next two steps consist in dividing
between
\begin{itemize}
  \item leading diagrams. Their $\varepsilon\to 0$ limit does not
  vanish.
  \item subleading diagrams. Their $\varepsilon\to 0$ limit
  vanishes.
\end{itemize}
Since (\ref{5.1}) is an infinite series, one needs on top
\begin{itemize}
  \item a uniform bound on the series.
  \item the leading diagrams at
  $\varepsilon=0$ sum to the time-dependent solution of the
  transport equation.
\end{itemize}

There are variations in the way how to arrange the perturbation
series. The expansion of $\langle U(t)^\ast a(y)^\ast
a(x)U(t)\rangle_\varepsilon$ yields standard Feynman diagrams.
Particle conservation is exploited more efficiently by
considering the hierarchy of $n$-particle Wigner functions,
$n=1,2,\ldots$ \cite{BCEP04,BCEP05,W80}. For a particle in a random potential the most
powerful method is to use directly (\ref{5.1}) \cite{EY00}.

A second basic choice is between $\mathbb{R}^3$ and $\mathbb{Z}^3$
as position space. $\mathbb{R}^3$ has the advantage that for phase
factors of the form $e^{-i k^2 t}$ the oscillatory integrals are
easily estimated. On the other hand the large $k$ behavior needs
extra efforts. For $\mathbb{Z}^3$ the free propagator is less
explicit, but momentum space is $\mathbb{T}^3$, hence bounded.

For weakly interacting systems the perturbation series diverges on
the kinetic time scale. This leaves one with analysing individual
diagrams. For a translation invariant Fermi gas all terms up to
order $(V_\mathrm{I})^2$ are studied \cite{HL97} and the
approximation $W(k,t)=W(k)+t\mathcal{C}(W)(k)$ is proved with
$\mathcal{C}$ of (\ref{4.5}) for $\theta=-1$ and integrations over
$\mathbb{T}^3$, $\omega(k)=\sum^3_{j=1}(1-\cos k^j)$. For bosons and
fermions moving in $\mathbb{R}^3$ the same result is obtained 
 for spatially inhomogeneous
initial data \cite{BCEP05}. The next set of results concerns the
analysis of a certain subclass of diagrams, e.g. all leading diagrams 
\cite{BCEP04} or the identity
permutation in (\ref{3.3}) which corresponds to Boltzmann
statistics \cite{BCEP07}. All these results strongly support the kinetic
picture. Spatial dimension has to be larger than three, which roughly
comes from conditions on oscillatory integrals as, e.g.,
\begin{equation}\label{5.2}
\int^\infty_0 dt|\int_{\mathbb{T}^d} dk e^{-i\omega(k)t}|<\infty\,.
\end{equation}

For the motion in a random potential the situation is more favorable.
Assume a Gaussian random potential. Then only even moments
contribute. For the $2n$-th term of the perturbation series this
yields a bound as
$\frac{1}{n!}\varepsilon^n(\varepsilon^{-1}t)^nc^n((2n)!/2^nn!)$, 
where the term in the second round brackets comes
from the number of pairings for the Gaussian random potential. Thus
the perturbation series has a finite radius of convergence, $t_0$, on the
kinetic scale. With this restriction the kinetic program can be
carried through. The spatially homogeneous case is dealt with in
\cite{S77}, which is substantially improved and extended in \cite{HLW92}, see also the
discussion in \cite{H93}. In 1997 L.~Erd\"{o}s and H.-T.~Yau
\cite{EY98} announced a truly remarkable break through. By a suitable
truncation of the perturbation series, the remainder being
controlled by unitarity, they succeed to prove the kinetic limit
for arbitrary $t_0$. With their technique other cases can be
handled as well, to list: low density scatterers \cite{EE05},
non-Gaussian but approximately independent random potentials \cite{ESY06}, tight binding
dynamics \cite{C06}, coupling to a phonon bath \cite{E02}, and mass
disordered lattice dynamics \cite{LS07}. Chen proves self-averaging
\cite{C07} and the link to spectral properties of $H$ \cite{C06}.

Staying in the context of the motion of an electron in a random
potential, one might wonder what happens beyond the kinetic time
scale. Mathematically this means to consider times of order
$\varepsilon^{-(1+\alpha)}t$ in microscopic units with $\alpha>0$ in
the limit $\varepsilon\to 0$. This implies very long kinetic times
at which the solution to the transport equation is already of the
form
\begin{equation}\label{5.3}
W(r,k,t)\cong \rho(r,t)\delta\big(E-\omega(k)\big)
\end{equation}
with $E$ determined by the initial conditions. The mass density $\rho$
has a slow variation and is governed by the diffusion equation
\begin{equation}\label{5.4}
\frac{\partial}{\partial t}\rho(r,t)=D_\textrm{kin}(E)\Delta_r
\rho(r,t)\,,
\end{equation}
where $D_\textrm{kin}(E)$ is the energy dependent diffusion coefficient of the kinetic
equation
\begin{equation}\label{5.5}
\frac{\partial}{\partial
t}W(r,k,t)+\nabla_k\omega(k)\cdot \nabla_r W(r,k,t)=
\mathcal{C}(W)(r,k,t)
\end{equation}
with $\mathcal{C}$ the collision operator from (\ref{4.4}). If $\alpha$ is large, the diffusion coefficient
in (\ref{5.4}) should be close to the diffusion coefficient as
computed for fixed small $\varepsilon$, which is definitely
different from $D_\textrm{kin}(E)$. Thus there must be a critical
$\alpha_c$ up to which (\ref{5.3}) together with (\ref{5.4}) is
valid. The theoretical prediction is $\alpha_c=1$. Erd\"{o}s,
Salmhofer, and Yau \cite{ESY07a,ESY07b} prove (\ref{5.3}), (\ref{5.4}) for
$0<\alpha<10^{-3}$.

\end{document}